\documentclass{ws-procs9x6}

\usepackage{dsfont}
\usepackage{graphicx}
\usepackage{appendix}
\usepackage{hyperref}
\usepackage{accents}
\usepackage[normalem]{ulem}
\usepackage{color}
\definecolor{blue}{rgb}{0,0,1}
\definecolor{red}{rgb}{1,0,0}

\usepackage[thmmarks]{ntheorem-hyper}
\theoremheaderfont{\bf}\theorembodyfont{\it}\theoremsymbol{}\theoremstyle{plain}\newtheorem{Thm}{Theorem}
\theoremheaderfont{\bf}\theorembodyfont{\rm}\theoremsymbol{\setlength{\unitlength}{1mm}\begin{picture}(2.5,2.5)\linethickness{2.5mm}\put(0,1.25){\line(1,0){2.5}}\end{picture}}\theoremstyle{plain}\newtheorem{Prf}{Proof}
\theoremheaderfont{\bf}\theorembodyfont{\it}\theoremsymbol{}\theoremstyle{plain}\newtheorem{Def}[Thm]{Definition}
\theoremheaderfont{\bf}\theorembodyfont{\sl}\theoremsymbol{}\theoremstyle{plain}\newtheorem{Rem}[Thm]{Remark}
\theoremheaderfont{\bf}\theorembodyfont{\it}\theoremsymbol{}\theoremstyle{plain}
\theoremheaderfont{\bf}\theorembodyfont{\sl}\theoremsymbol{}\theoremstyle{plain}
\theoremheaderfont{\bf}\theorembodyfont{\it}\theoremsymbol{}\theoremstyle{plain}\newtheorem{Col}[Thm]{Corollary}
\newcommand{\Aut}{\mathop{\mathrm{Aut}}\nolimits}
\newcommand{\InAut}{\mathop{\mathrm{InAut}}\nolimits}

\newcommand{\Real}{\mathop{\mathrm{Re}}\nolimits}

\newcommand{\1}{\mathds{1}}
\newcommand{\R}{\mathbb{R}}
\newcommand{\C}{\mathbb{C}}
\newcommand{\Z}{\mathbb{Z}}
\newcommand{\Zt}{\mathbb{Z}_{2}}
\newcommand{\Zs}{\mathbb{Z}\times\mathbb{Z}}
\newcommand{\ev}{\mathrm{ev}}
\newcommand{\od}{\mathrm{od}}
\newcommand{\GL}{\mathrm{GL}}

\newcommand{\Lin}{\mathrm{Lin}}
\newcommand{\Ann}{\mathrm{Ann}}
\newcommand{\triangled}{\mathbin{\rotatebox[origin=c]{180}{$\triangle$}}}

\newcommand\tbar[1]{\accentset{\rule{.4em}{.8pt}}{#1}}
\newlength{\figuresize}
\newlength{\figurevgap}
\newlength{\figurehgap}
\begin{document}

\title{A NATURAL EXTENSION OF THE CONFORMAL LORENTZ GROUP IN A FIELD THEORY CONTEXT}

\author{Andr\'as L\'aszl\'o}

\address{Wigner Research Centre for Physics of the Hungarian Academy of Sciences,\\
RMI, Department of High Energy Physics\\
Budapest 1121, P.O.Box 49, H-1525, Hungary\\
\texttt{laszlo.andras@wigner.mta.hu}}

\begin{abstract}
In this paper a finite dimensional unital associative algebra is presented, 
and its group of algebra automorphisms is detailed. The studied algebra 
can physically be understood as the creation operator algebra in a formal 
quantum field theory at fixed momentum for a spin 1/2 particle along with 
its antiparticle. It is shown that the essential part of the corresponding 
automorphism group can naturally be related to the conformal Lorentz group. 
In addition, the non-semisimple part of the automorphism group can be 
understood as ``dressing'' of the pure one-particle states. The studied 
mathematical structure may help in constructing quantum field theories in a 
non-perturbative manner. In addition, it provides a simple example of 
circumventing Coleman-Mandula theorem using non-semisimple groups, 
without SUSY.
\end{abstract}


\vspace{5mm}

\bodymatter

\section{Introduction}
\label{introduction}

It is a well known fact in quantum field theory (QFT) that any kind of unification 
of internal (gauge) symmetries with spacetime symmetries is not evident at all. The 
celebrated Coleman-Mandula theorem and its various versions \cite{weinberg2000} 
prohibit the most simple unification scenarios: under quite generic conditions 
one cannot find a larger symmetry group for a QFT model, which is composed as 
the semi-direct product of the Poincar\'e group and the internal symmetry group. 
One of the important assumptions of the pertinent no-go theorem is that the Lie 
group of internal symmetries is restricted to products of copies of 
$\mathrm{U}(1)$ group and compact semisimple Lie groups. This means that the 
total symmetries of matter fields at a fixed spacetime point ---or equivalently, at a fixed point 
of momentum space--- is a product of copies of $\mathrm{U}(1)$ group and a semisimple Lie group.

Our aim is to show a possible, physically natural mathematical example, 
when the total symmetry group of matter fields at a spacetime point 
---or equivalently, at a fixed point of momentum space--- is some nontrivial 
extension of the (conformal) Lorentz group, or more precisely, its covering 
group. The pertinent extended group can be viewed as the automorphisms of 
the creation operator algebra for a spin 1/2 particle along with its antiparticle. 
The non-semisimpleness of the discussed group gives possibility to nontrivially 
extend the group of spacetime symmetries. The extended part becomes an idempotent 
normal subgroup, which can be regarded as ``dressing transformations'' of pure 
one-particle states in a formal QFT context at a fixed momentum. The presented 
algebraic construction may also help to improve mathematical formulation of 
QFTs: it shows a possibility to avoid building the theory based on first 
constructing the one-particle theory and then constructing the corresponding 
multi-particle model, \emph{a posteriori}. That is because the ``dressing transformations'' mix 
the pure $p$-particle,$q$-antiparticle states with each-other to some extent, 
putting the possible particle combination states in a unified multiplet.

\section{Preliminaries: automorphisms of a Grassmann algebra}
\label{preliminaries}

First we review the properties of the automorphism group of finite dimensional 
Grassmann algebras as prototype problem \cite{berezin1967,djokovic1978,bavula2007,bavula2009}. 
Physically, an $n$-generator Grassmann algebra can be thought of as the 
algebra of creation operator polynomials in a formal QFT at fixed momentum 
of a fermion particle with $n$ internal degrees of freedom.

\begin{Def} (Grassmann algebra, canonical generator system)
A finite dimensional complex associative algebra $G$ with unit is called a 
\textbf{Grassmann algebra} if there exists a minimal generating system 
$(e_{1},\dots,e_{n})$ of $G$ such that
\begin{eqnarray}
e_{i}e_{j} + e_{j}e_{i} \;\,=0 & \qquad (i,j\in\{1,\dots,n\}) & \mathrm{and} \cr
e_{i_{1}}e_{i_{2}} \dots e_{i_{k}} & \qquad (1\leq i_{1}<i_{2}<\dots<i_{k}\leq n,\; 0\leq k\leq n) \cr
 & \qquad \mathrm{are}\;\mathrm{linearly}\;\mathrm{independent}.
\label{grassmanngeneratorrelations}
\end{eqnarray}
Such a minimal generating system shall be referred to as \textbf{canonical generator system}.
\end{Def}

We will now explore the basic properties of the group of automorphisms $\Aut(G)$ of $G$. 
These are the $G\rightarrow G$ invertible complex-linear transformations, which 
preserve the algebraic product on $G$.

\begin{Rem} The following basic properties are well known\cite{berezin1967,djokovic1978,bavula2007,bavula2009,bourbaki1989}.

(i) Grassmann algebras do exist: the exterior algebra $\Lambda(V)$ of a 
finite dimensional complex vector space $V$ is a Grassmann algebra, where a basis of $V$ 
are canonical generators. In fact, all Grassmann algebras are (not naturally) isomorphic to some 
exterior algebra $\Lambda(V)$.

(ii) If $(e_{1},\dots,e_{n})$ is a canonical generator system and $\alpha\in\Aut(G)$, 
then $(\alpha(e_{1}),\dots,\alpha(e_{n}))$ are also. If, in addition, $(e_{1}',\dots,e_{n}')$ is 
an other system of canonical generators, then a mapping 
$e_{i}\mapsto \alpha(e_{i}):=e_{i}'$ ($i\in\{1,\dots,n\}$) uniquely determines an automorphism $\alpha\in\Aut(G)$. 
That is, automorphisms can uniquely be characterized by their action on an arbitrarily chosen canonical generator system.

(iii) Given a chosen system of canonical generators, the linear subspace of the 
pure $k$-th order polynomials of them are called the space of $k$-forms, and are 
denoted by $\Lambda_{k}$. The linear subspace of the pure even / odd polynomials 
of them are called the space of even / odd forms, and are denoted by $\Lambda_{\ev}$ / $\Lambda_{\od}$. 
As such, one has $G=\mathop{\oplus}\limits_{k=0}^{n} \Lambda_{k}$ and $G=\Lambda_{\ev}\oplus\Lambda_{\od}$. 
These splittings of $G$ are referred to as $\Z$ and $\Zt$-grading, respectively.

(iv) Let the unity be denoted by $\1$ and its complex linear span by $B$. Let 
$M:=\mathop{\oplus}\limits_{k=1}^{n} \Lambda_{k}$ be the linear subspace of at least $1$-forms 
and observe that it is the maximal ideal of $G$, and therefore is an 
$\Aut(G)$-invariant subspace. Because of that, we have the 
$\Aut(G)$-invariant splitting $G=B\oplus M$ with corresponding $\Aut(G)$-invariant 
complementing projection operators $I-m$ and $m$. Because of $\Aut(G)$-invariance of 
unity, $I-m=\1 b$ can be written with uniquely determined $b:\,G\rightarrow\C$, $\Aut(G)$-invariant map, picking out the scalar component.

(v) Since $M$ is $\Aut(G)$-invariant, all its powers $M^{l}=\mathop{\oplus}\limits_{k=l}^{n} \Lambda_{k}$ are 
$\Aut(G)$-invariant ($l\in\{1,\dots,n\}$).

(vi) The center of $G$, denoted by $Z(G)$, consists of all elements commuting 
with $G$, and they form an $\Aut(G)$-invariant subspace.
\end{Rem}

\begin{Thm} (D.~Z.~Djokovic \cite{djokovic1978})
Let us define the following subgroups of $\Aut(G)$, given a canonical generator system $(e_{1},\dots,e_{n})$ of $G$.

(i) Let $\Aut_{\Z}(G)$ be the $\Z$-grading preserving automorphisms. These are of the form 
$e_{i}\mapsto\sum_{j=1}^{n}\alpha_{ij}e_{j}$ ($i\in\{1,\dots,n\}$) with 
$\left(\alpha_{ij}\right)_{i,j\in\{1,\dots,n\}}\in\GL\left(\C^{n}\right)$.

(ii) Let $N_{\ev}$ be the $\Zt$-grading preserving automorphisms acting on $M^{1}/M^{2}$ factor space 
as unity. These are of the form $e_{i}\mapsto e_{i}+b_{i}$ ($i\in\{1,\dots,n\}$) with 
$b_{i}\in M^{3}\cap \Lambda_{\od}$.

(iii) Let $\InAut(G)$ be the inner automorphisms, i.e.\ the ones of the form $\exp(a)(\cdot)\exp(a)^{-1}$ (with some $a\in G$). These are of the form 
$e_{i}\mapsto e_{i}+\left[a,e_{i}\right]$ ($i\in\{1,\dots,n\}$) with some $a\in G$.

With these, the semi-direct product splitting
\begin{eqnarray}
\Aut(G) = \InAut(G) \rtimes N_{\ev} \rtimes \Aut_{\Z}(G)
\end{eqnarray}
holds.
\end{Thm}

\begin{Col}
As a consequence, we have that the $\Lambda_{k}$ subspaces of $k$-forms are not 
$\Aut(G)$-invariant. In fact, the list of indecomposable $\Aut(G)$-invariant 
subspaces are the followings: $B$, $M^{l}$ ($l\in\{1,\dots,n\}$), $M^{l}\cap Z(G)$ ($l\in\{2,\dots,n\}$).
\end{Col}

Physicswise, this means that if we think of $G$ as the algebra of creation 
operator polynomials in a formal QFT of an $n$ internal degrees of freedom 
fermion particle at fixed momentum, and we assume that the full $\Aut(G)$ acts 
on this algebra as symmetry group, then it becomes a unified multiplet. 
Particularly, the only $\Aut(G)$-invariant decomposition is $G=B\oplus M$, 
i.e.\ the splitting to $0$-particle and to at-least-$1$-particle states. The 
reason is that the normal subgroup $N:=\InAut(G) \rtimes N_{\ev}$ mixes higher 
particle content to lower particle states. This motivates to call $N$ the 
\emph{\textbf{dressing transformations}}, being an idempotent normal subgroup of $\Aut(G)$.

\section{Spin algebra and its automorphisms}
\label{spinalgebra}

Motivated by the above findings, we define a physically more relevant setting.

\begin{Def} (${}^{+}$-algebra) A finite dimensional complex associative algebra $A$ with unit 
shall be called a \textbf{
${}^{+}$-algebra} if it is equipped with a conjugate-linear involution satisfying 
$(xy)^{+}=x^{+}y^{+}$ for all $x,y\in A$.
\end{Def}

It is important that in the above definition the ${}^{+}$-adjoint does not reverse 
the order of products, i.e.\ it is slightly different than that of a usual ${}^{*}$-algebra. 
This will, physically, model the charge conjugation in our construction.

\begin{Def} (spin algebra) A finite dimensional complex associative ${}^{+}$-algebra $A$ with unit 
shall be called a \textbf{spin algebra} if there exists a minimal generating system 
$(e_{1},e_{2},e_{3},e_{4})$ of $A$ such that
\begin{eqnarray}
e_{i}e_{j} + e_{j}e_{i} \;\,=0 & \qquad (i,j\in\{1,2\}\;\mathrm{or}\;i,j\in\{3,4\}) & \mathrm{and} \cr
e_{i}e_{j} - e_{j}e_{i} \;\,=0 & \qquad (i\in\{1,2\}\;\mathrm{and}\;j\in\{3,4\}) & \mathrm{and} \cr
e_{3}=e_{1}^{+},\; e_{4}=e_{2}^{+} & & \mathrm{and} \cr
e_{i_{1}}e_{i_{2}} \dots e_{i_{k}} & \qquad (1\leq i_{1}<i_{2}<\dots<i_{k}\leq 4,\; 0\leq k\leq 4) & \cr
 & \qquad \mathrm{are}\;\mathrm{linearly}\;\mathrm{independent}. &
\label{spinalgebrageneratorrelations}
\end{eqnarray}
Such a minimal generating system shall be referred to as \textbf{canonical generator system}, and 
the notation $n:=4$ will occasionally be used.
\end{Def}

Physicswise, a spin algebra can be thought of as the algebra of creation 
operator polynomials in a formal QFT of a $2$ internal degree of freedom 
fermion particle along with its antiparticle, at fixed momentum. It is important to note 
that in this construction the annihilation operators of particles are not yet 
identified with the creation operator of antiparticles, and therefore a spin 
algebra is very different than that of a canonical anticommutation relation (CAR) algebra, 
also referred to as mixed exterior algebras\cite{vanstone1984}. Here, the ${}^{+}$-adjoining 
only models the charge conjugation operator, exchanging particle and antiparticle 
creation operators in a conjugate-linear way. 
We will now explore the basic properties of the group of automorphisms $\Aut(A)$ of $A$. 
These are the $A\rightarrow A$ invertible complex-linear transformations, which 
preserve the algebraic product on $A$ along with the ${}^{+}$-adjoining.

\begin{Rem} Similar properties hold as for a Grassmann algebra.

(i) Spin algebras do exist: if $S^{*}$ is a complex $2$ dimensional vector space, 
called to be the cospinor space, and we take $\Lambda(\bar{S}^{*})\otimes\Lambda(S^{*})$, 
then it naturally becomes a spin algebra. In fact, all spin algebras are (not naturally) isomorphic to this. 
(The notation $\bar{(\cdot)}$ means complex conjugation.)

(ii) If $(e_{1},\dots,e_{n})$ is a canonical generator system and $\alpha\in\Aut(A)$, 
then $(\alpha(e_{1}),\dots,\alpha(e_{n}))$ are also. If, in addition, $(e_{1}',\dots,e_{n}')$ is 
an other system of canonical generators, then a mapping 
$e_{i}\mapsto \alpha(e_{i}):=e_{i}'$ ($i\in\{1,\dots,n\}$) uniquely determines an automorphism $\alpha\in\Aut(A)$. 
That is, automorphisms can uniquely be characterized by their action on an arbitrarily chosen canonical generator system.

(iii) Given a chosen system of canonical generators, the linear subspace of the 
pure polynomials of $p$ pieces of $\{e_{1},e_{2}\}$ and $q$ pieces of 
of $\{e_{1}^{+},e_{2}^{+}\}$ are called the space of $p,q$-forms, and are 
denoted by $\Lambda_{\bar{p},q}$. The subspace of those polynomials for which $p+q=k$ holds 
are called the space of $k$-forms. The linear subspace of the pure even / odd $\Lambda_{k}$-s 
are called the space of even / odd forms, and are denoted by $\Lambda_{\ev}$ / $\Lambda_{\od}$. 
As such, one has $A=\mathop{\oplus}\limits_{p,q=0}^{2}\Lambda_{\bar{p},q}$, $A=\mathop{\oplus}\limits_{k=0}^{4} \Lambda_{k}$ and $A=\Lambda_{\ev}\oplus\Lambda_{\od}$. 
These splittings of $A$ are referred to as the $\Zs$, $\Z$ and $\Zt$-grading, respectively.

(iv) Let the unity be denoted by $\1$ and its complex linear span by $B$. Let 
$M:=\mathop{\oplus}\limits_{k=1}^{n} \Lambda_{k}$ be the linear subspace of at least $1$-forms 
and observe that it is the maximal ideal of $A$, and therefore is an 
$\Aut(A)$-invariant subspace. Because of that, we have the 
$\Aut(A)$-invariant splitting $A=B\oplus M$ with corresponding $\Aut(A)$-invariant 
complementing projection operators $I-m$ and $m$. Because of $\Aut(A)$-invariance of 
unity, $I-m=\1 b$ can be written with uniquely determined $b:\,A\rightarrow\C$, $\Aut(A)$-invariant map.

(v) Since $M$ is $\Aut(A)$-invariant, all its powers $M^{l}=\mathop{\oplus}\limits_{k=l}^{n} \Lambda_{k}$ are 
$\Aut(A)$-invariant ($l\in\{1,\dots,n\}$).

(vi) The center of $A$, denoted by $Z(A)$, consists of all elements commuting 
with $A$, and they form an $\Aut(A)$-invariant subspace. In fact, $Z(A)=\Lambda_{\bar{0}0}\oplus \Lambda_{\bar{2}0}\oplus \Lambda_{\bar{0}2}\oplus \Lambda_{\bar{2}2}$.
\end{Rem}

\begin{Thm} (A.~L\'aszl\'o \cite{laszlo2015})
Let us define the following subgroups of $\Aut(A)$, given a canonical generator system $(e_{1},e_{2},e_{1}^{+},e_{2}^{+})$ of $A$.

(i) Let $\Aut_{\Zs}(A)$ be the $\Zs$-grading preserving automorphisms. These are of the form 
$e_{i}\mapsto\sum_{j=1}^{2}\alpha_{ij}e_{j}$ and $e_{i}^{+}\mapsto\sum_{j=1}^{2}\bar{\alpha}_{ij}e_{j}^{+}$ ($i\in\{1,2\}$) with 
$\left(\alpha_{ij}\right)_{i,j\in\{1,2\}}\in\GL\left(\C^{2}\right)$. Here, as before, 
$\bar{(\cdot)}$ denotes complex conjugation.

(ii) Let $\mathcal{J}:=\{I,J\}$ be the two element group of automorphisms, $I$ being 
the identity and $J$ being the involutive complex-linear operator of particle-antiparticle label exchanging: 
$e_{1}\mapsto e_{3}$, $e_{2}\mapsto e_{4}$, $e_{3}\mapsto e_{1}$, $e_{4}\mapsto e_{2}$.

(iii) Let $\tilde{N}_{\ev}$ be a subgroup of the $\Zt$-grading preserving 
automorphisms defined by the relations 
$e_{i}\mapsto e_{i}+b_{i}$ and $e_{i}^{+}\mapsto e_{i}^{+}+b_{i}^{+}$ ($i\in\{1,2\}$) with 
uniquely determined parameters $b_{i}\in \Lambda_{\bar{1}2}$.

(iv) Let $\InAut(A)$ be the inner automorphisms, i.e.\ the ones of the form $\exp(a)(\cdot)\exp(a)^{-1}$ (with some $a\in \Real(A)$). These are of the form 
$e_{i}\mapsto e_{i}+\left[a,e_{i}\right]+\frac{1}{2}\left[a,\left[a,e_{i}\right]]\right]$ ($i\in\{1,\dots,n\}$) with uniquely determined 
parameter $a\in\Real\left(\Lambda_{\bar{1}0}\oplus\Lambda_{\bar{0}1}\oplus\Lambda_{\bar{1}1}\oplus\Lambda_{\bar{2}1}\oplus\Lambda_{\bar{1}2}\right)\subset \Real(A)$.

With these, the semi-direct product splitting
\begin{eqnarray}
\Aut(A) = \InAut(A) \rtimes \tilde{N}_{\ev} \rtimes \Aut_{\Zs}(A) \rtimes \mathcal{J}
\end{eqnarray}
holds.
\end{Thm}
\begin{Prf} The proof is based on the fact that $\Aut(A)$ elements are the linear 
transformations preserving the canonical generator relations. Although 
the proof is theoretically not complicated, it is a quite extended calculation, and will be published 
in a more detailed paper \cite{laszlo2015}.
\end{Prf}

\begin{figure}[!ht]
\begin{center}
\includegraphics[height=\figuresize]{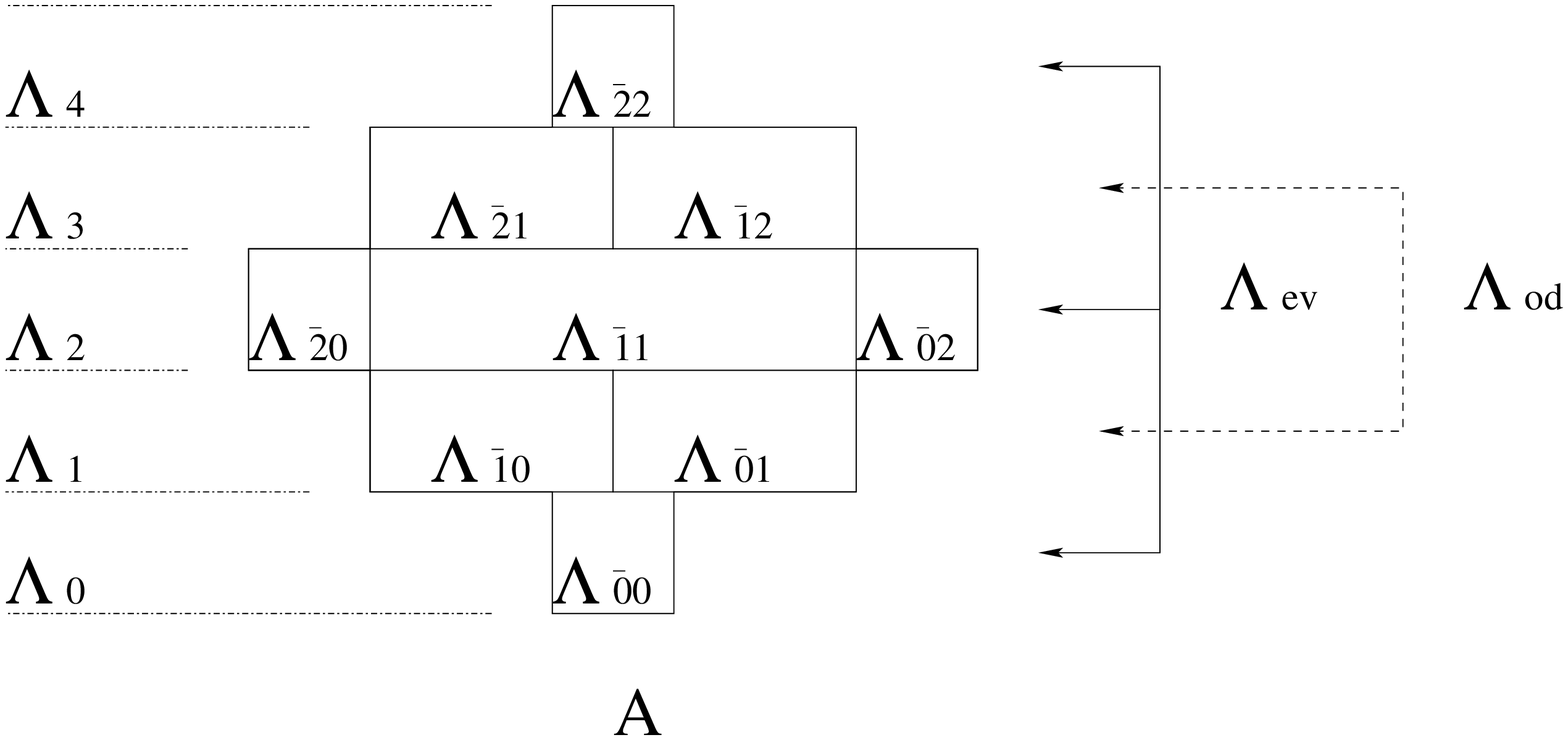}\hspace*{\figurehgap}\hspace*{\figurehgap}\includegraphics[height=\figuresize]{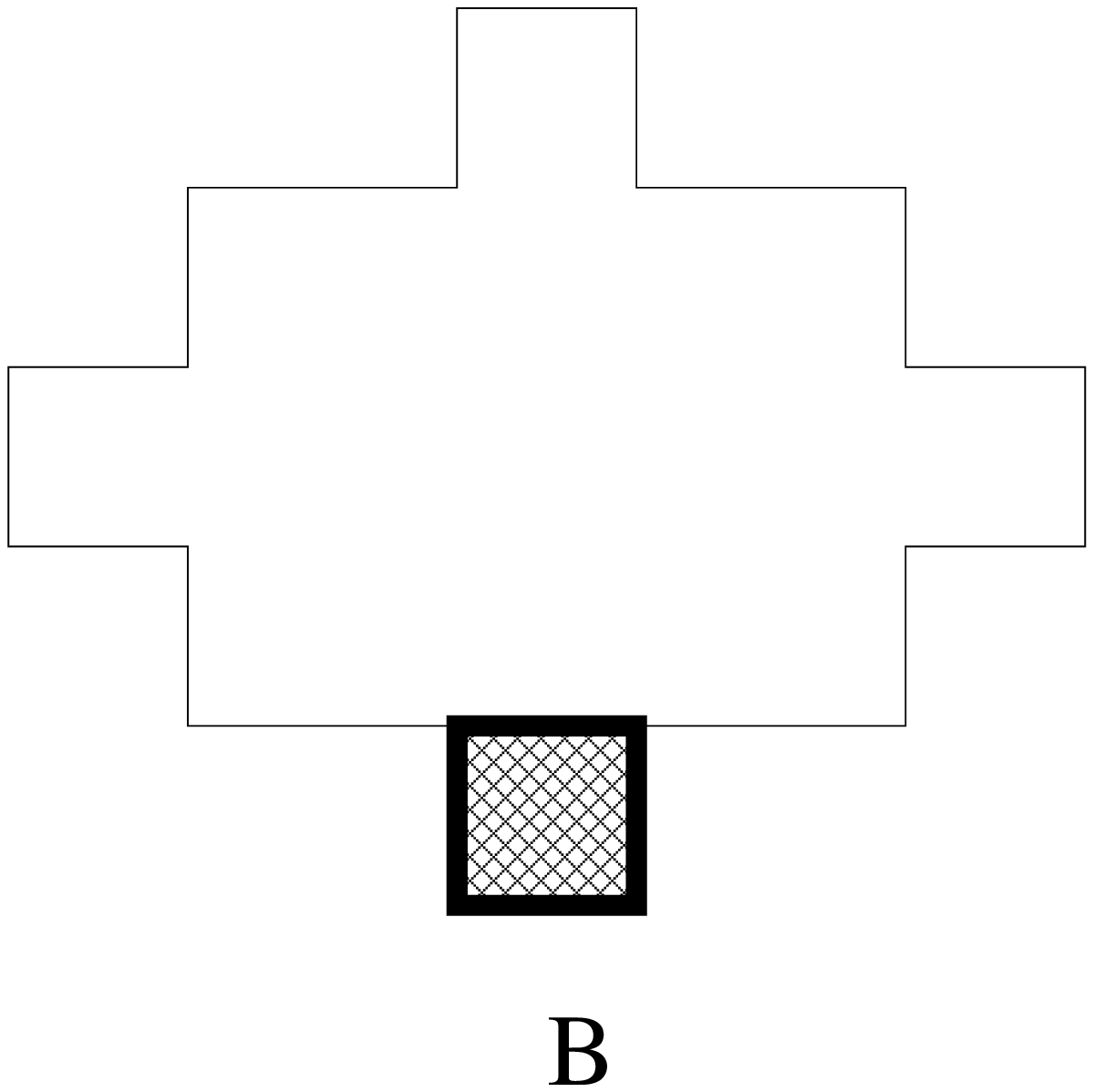}\\
\vspace*{\figurevgap}
\includegraphics[height=\figuresize]{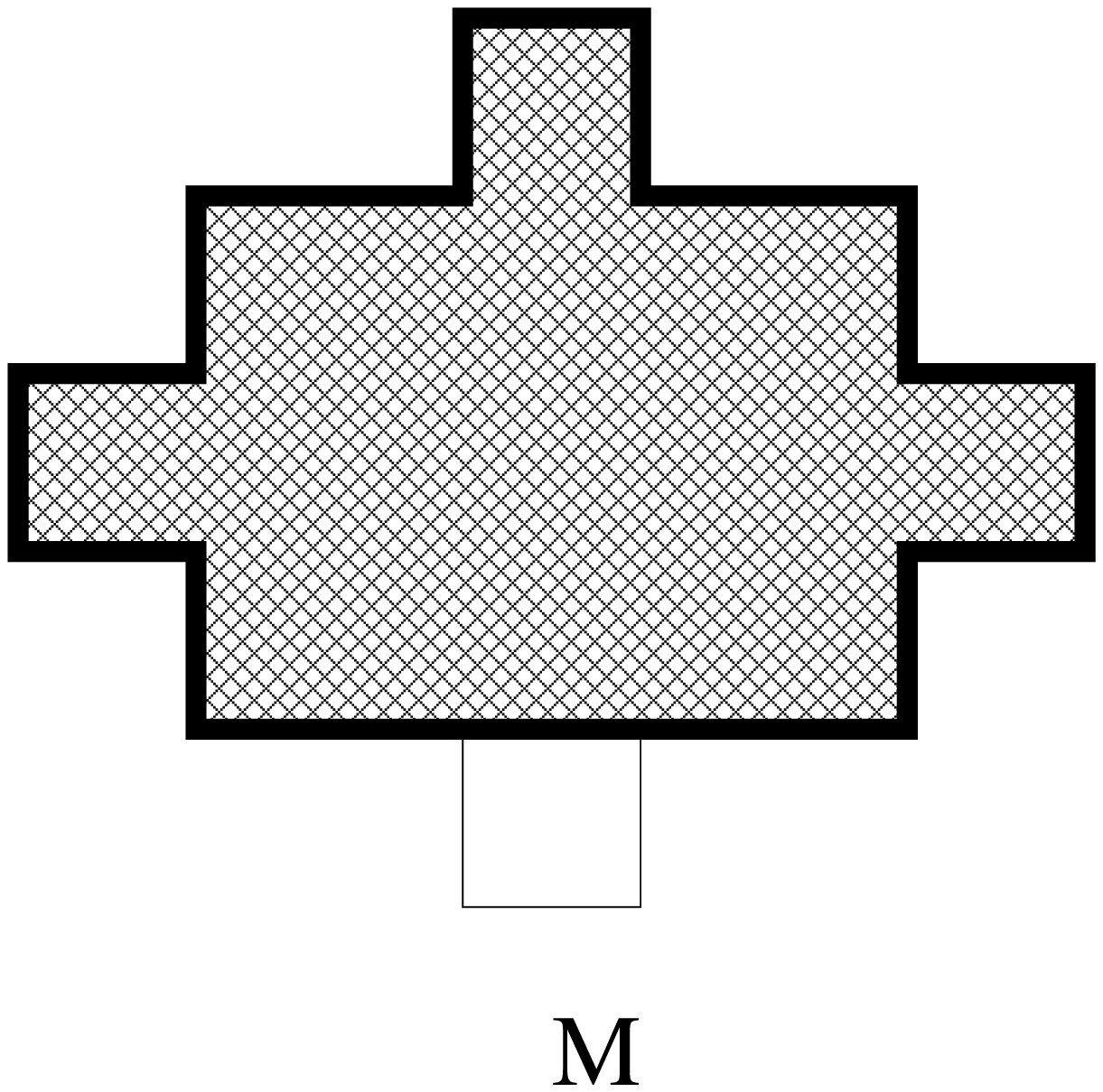}\hspace*{\figurehgap}\includegraphics[height=\figuresize]{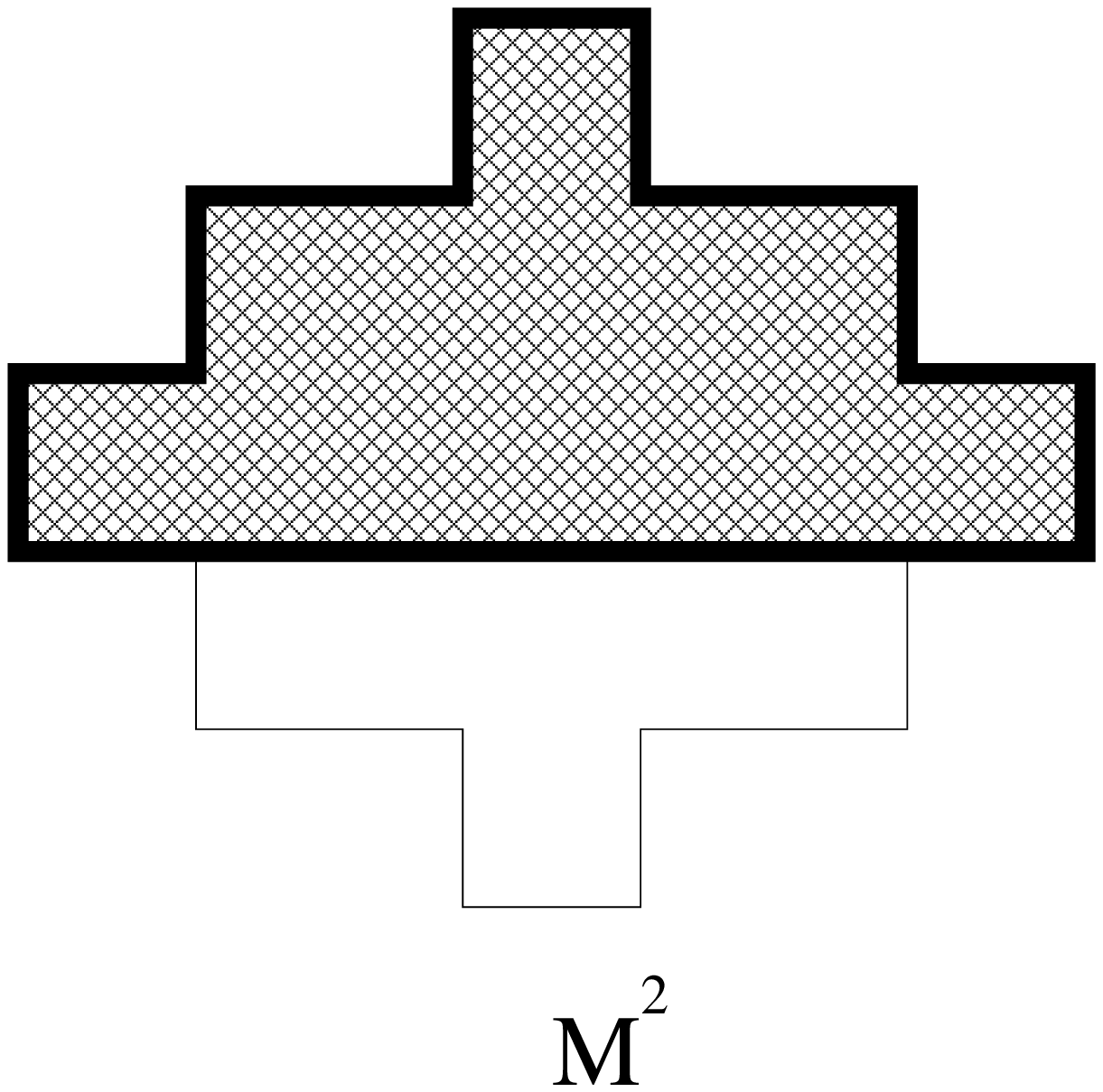}\hspace*{\figurehgap}\includegraphics[height=\figuresize]{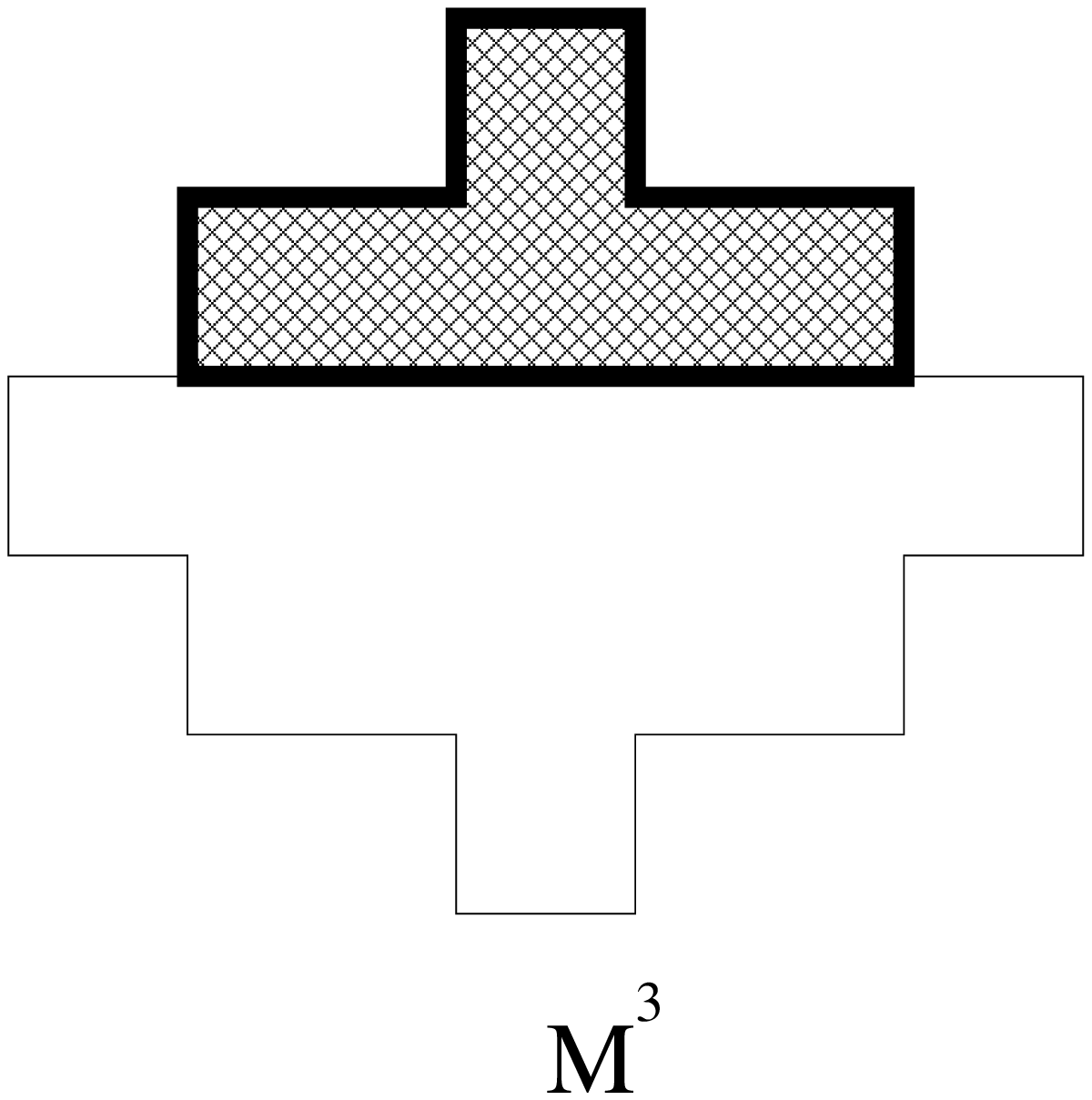}\\
\vspace*{\figurevgap}
\includegraphics[height=\figuresize]{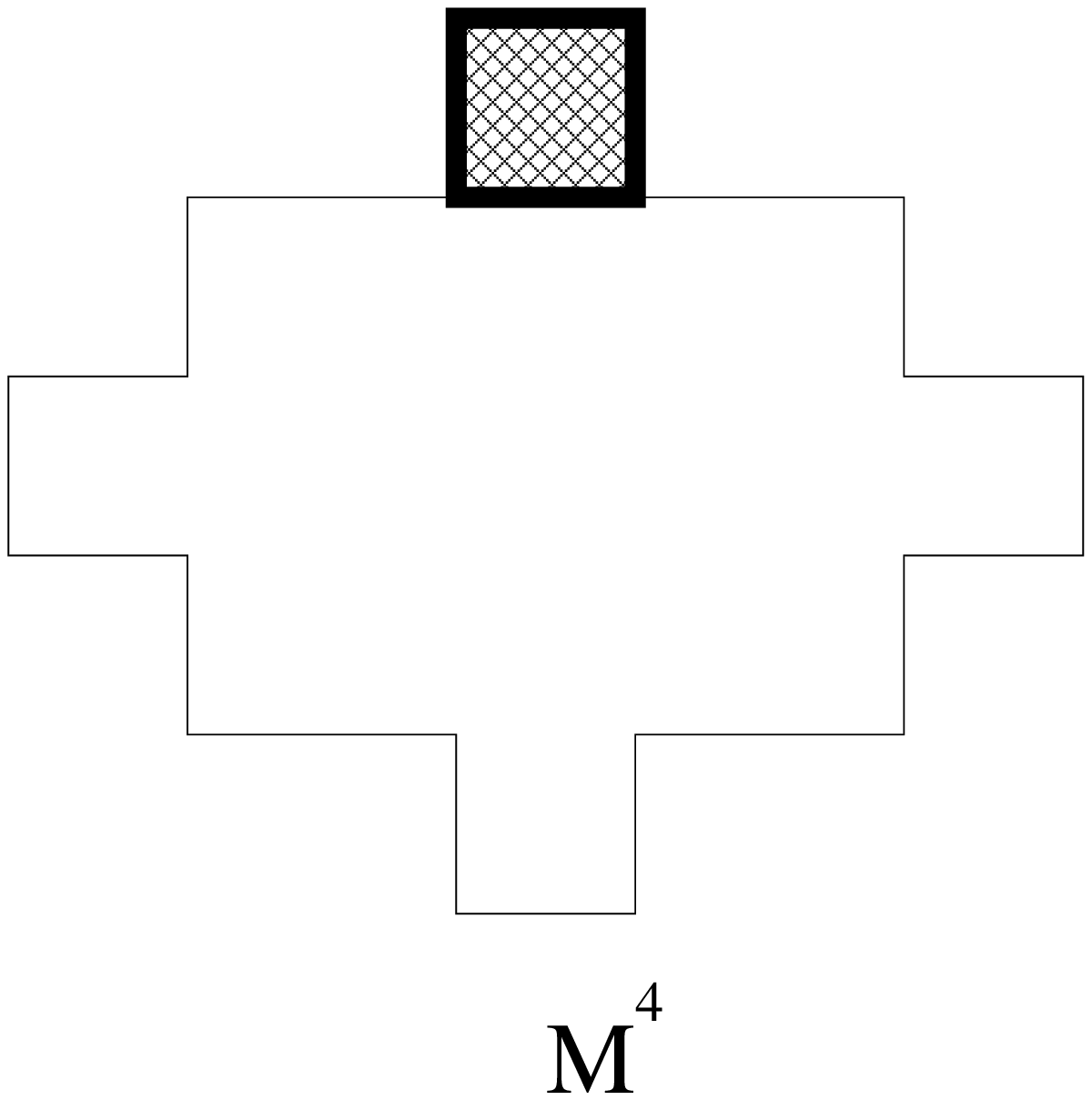}\hspace*{\figurehgap}\includegraphics[height=\figuresize]{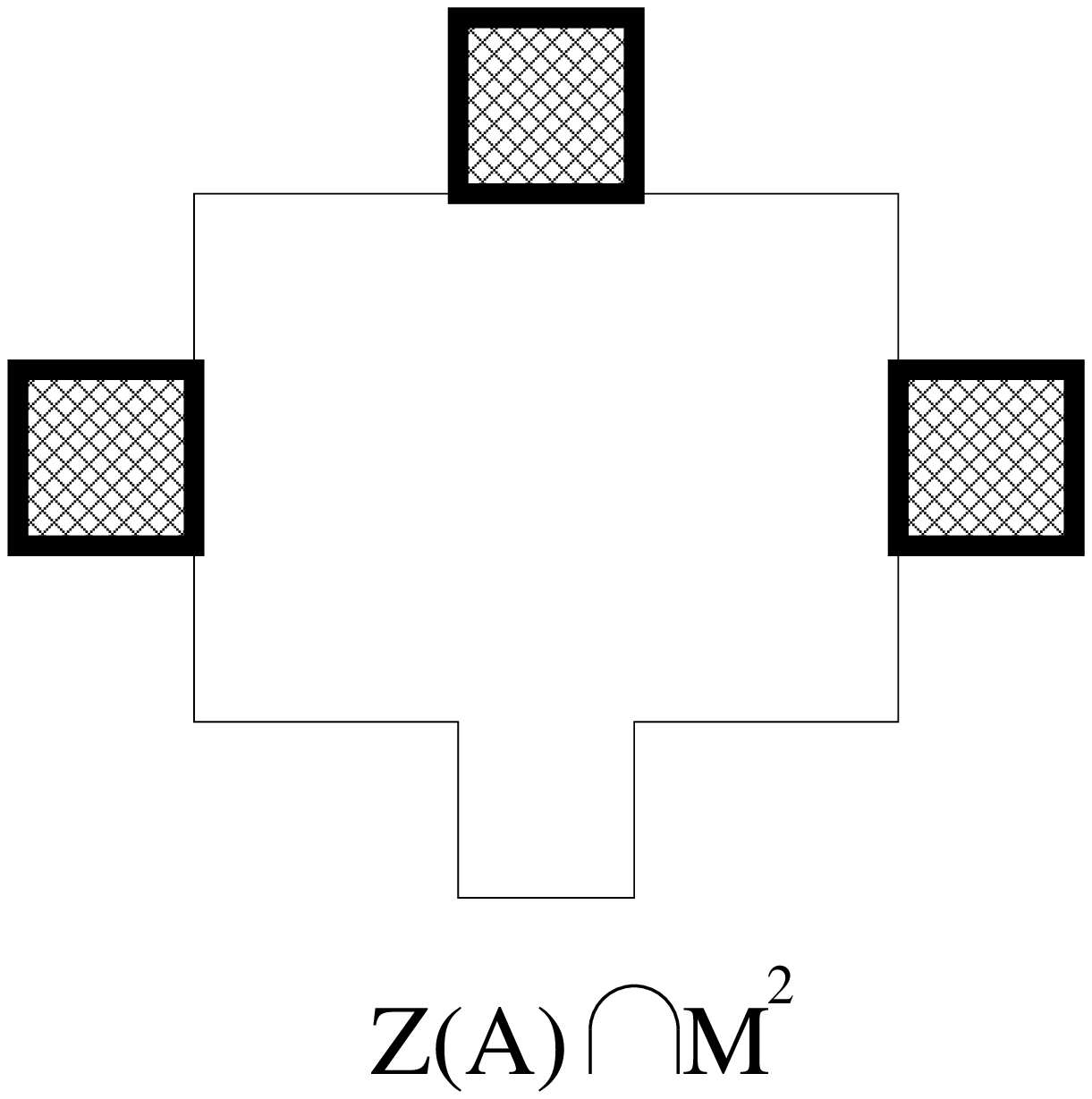}\hspace*{\figurehgap}\includegraphics[height=\figuresize]{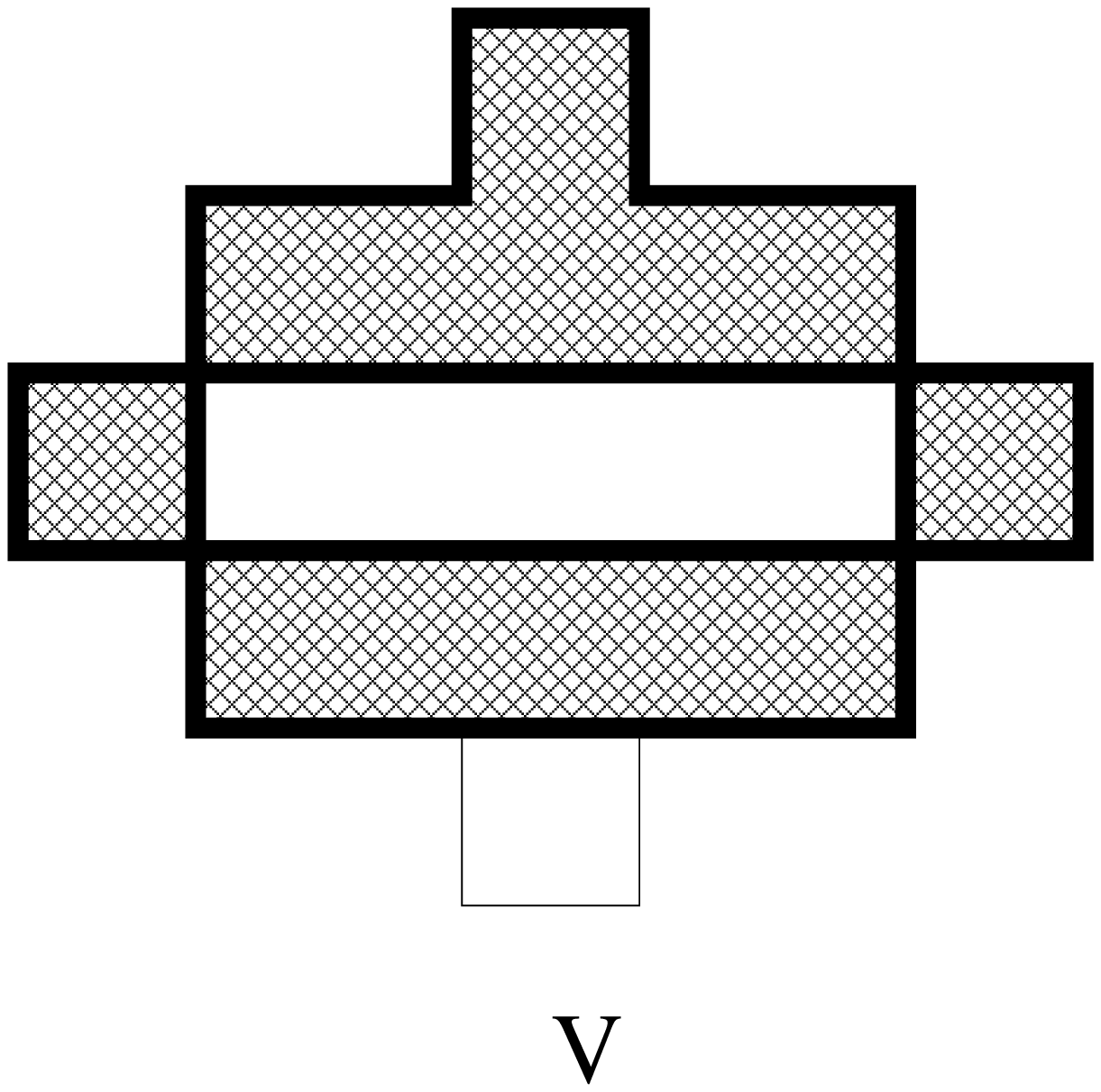}\\
\vspace*{\figurevgap}
\includegraphics[height=\figuresize]{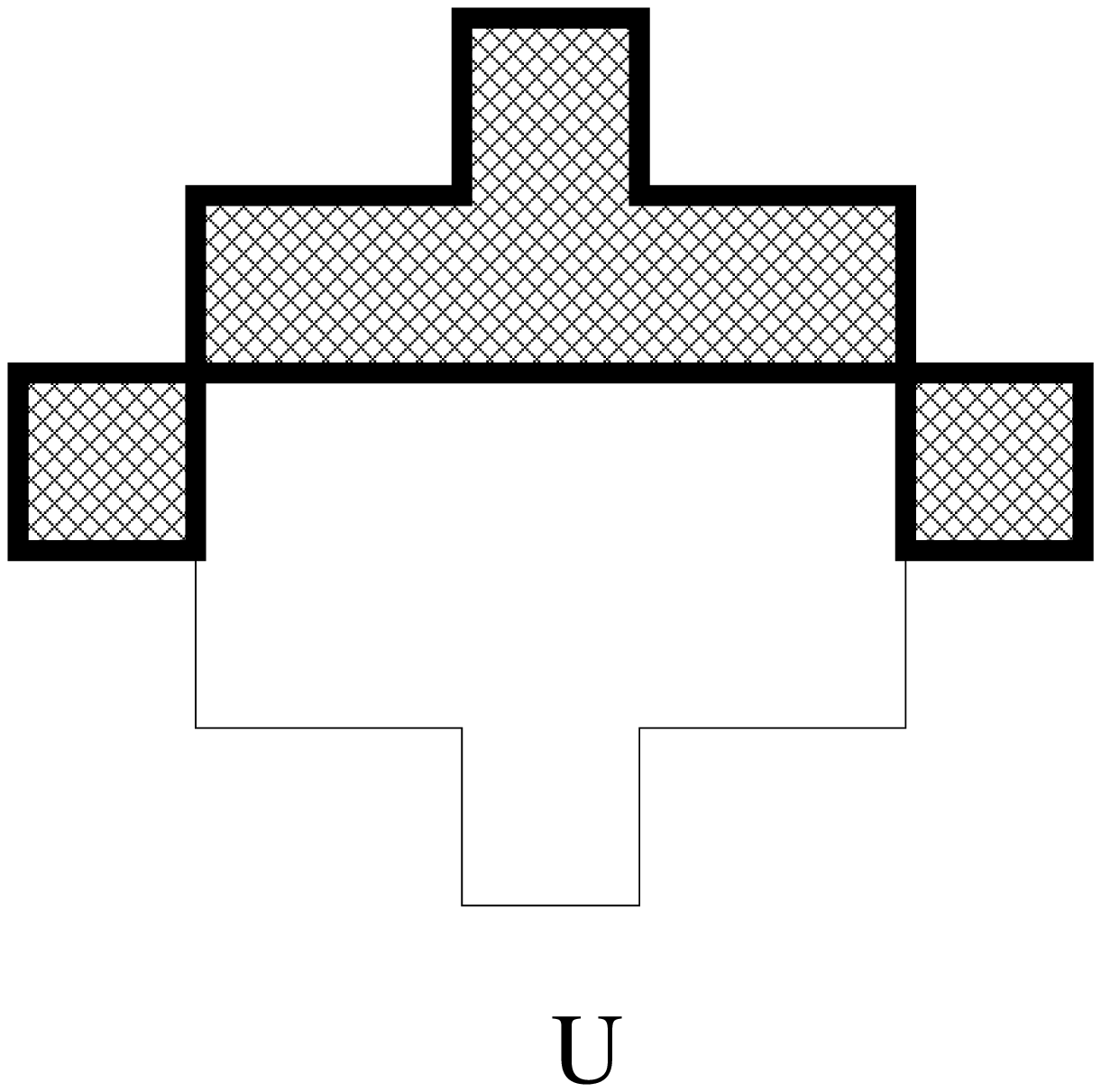}\hspace*{\figurehgap}\includegraphics[height=\figuresize]{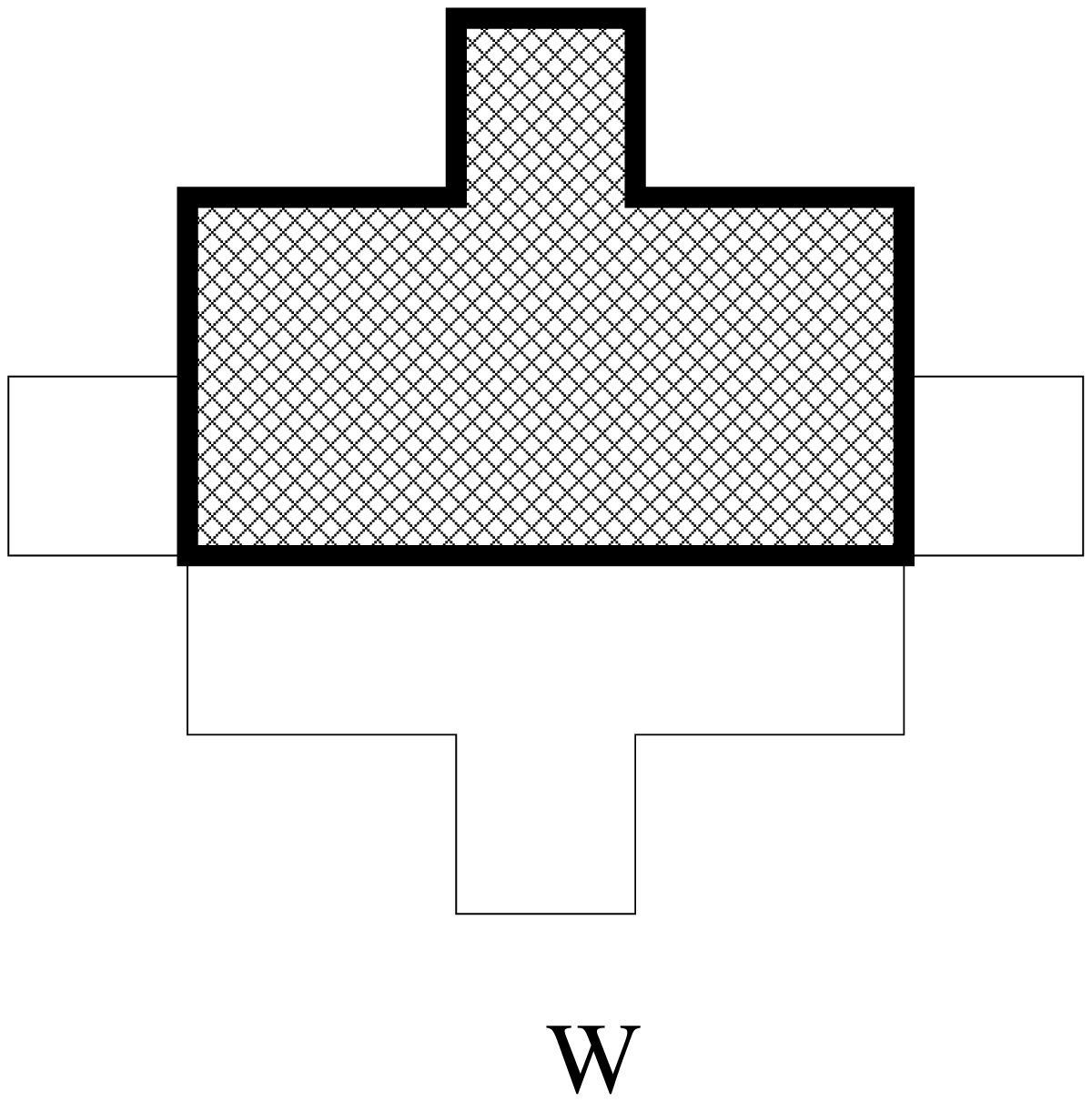}\\
\end{center}
\caption{Top left panel: illustration of $\Zs$, $\Z$ and $\Zt$-graded structure of the 
spin algebra $A$. Other panels: illustration of $\Aut(A)$-invariant indecomposable 
subspaces of $A$. One unit box depicts one complex dimension on all panels, shaded 
framed regions depict the pertinent indecomposable $\Aut(A)$-invariant subspaces.}
\label{figspininvsubsp}
\end{figure}

\begin{Col}
As a consequence, we have that the $\Lambda_{\bar{p}q}$ subspaces of $p,q$-forms are not 
$\Aut(A)$-invariant. In fact, the list of indecomposable $\Aut(A)$-invariant 
subspaces are the followings: $B$, $M^{l}$ ($l\in\{1,2,3,4\}$), $M^{2}\cap Z(A)$, $V:=\Lambda_{\bar{1}0}\oplus\Lambda_{\bar{0}1}\oplus\Lambda_{\bar{2}0}\oplus\Lambda_{\bar{0}2}\oplus\Lambda_{\bar{2}1}\oplus\Lambda_{\bar{1}2}\oplus\Lambda_{\bar{2}2}$, 
$U:=\Lambda_{\bar{2}0}\oplus\Lambda_{\bar{0}2}\oplus\Lambda_{\bar{2}1}\oplus\Lambda_{\bar{1}2}\oplus\Lambda_{\bar{2}2}$, $W:=\Lambda_{\bar{1}1}\oplus\Lambda_{\bar{2}1}\oplus\Lambda_{\bar{1}2}\oplus\Lambda_{\bar{2}2}$. 
These are illustrated in Fig.\ref{figspininvsubsp}.
\end{Col}

Physicswise, this result means that if we think of $A$ as the algebra of creation 
operator polynomials in a formal QFT of a $2$ internal degrees of freedom 
fermion particle along with its antiparticle, at fixed momentum, and we assume that the full $\Aut(A)$ acts 
on this algebra as symmetry group, then it becomes a unified multiplet. 
Particularly, the only $\Aut(A)$-invariant decomposition is $A=B\oplus M$, 
i.e.\ the splitting to $0$-particle and to at-least-$1$-particle states. The 
reason is that the normal subgroup $N:=\InAut(A) \rtimes \tilde{N}_{\ev}$ mixes higher 
particle content to lower particle states. This motivates to call $N$ the 
\emph{\textbf{dressing transformations}}, being an idempotent normal subgroup of $\Aut(A)$.

When dealing with the $\Zs$-grading preserving part of $\Aut(A)$, i.e.\ the 
action of the subgroup $\Aut_{\Zs}(A)\equiv\GL(\C^{2})\equiv \mathrm{D}(1)\times\mathrm{U}(1)\times\mathrm{SL}(\C^{2})$, a fixed $\Zs$ grading 
may be used, and therefore the whole formalism can be thought of as an ordinary 
two-spinor calculus \cite{penrose1984,wald1984} on the tensor algebra of 
cospinors and complex conjugate cospinors. Using this, it is immediately seen, 
that $\Aut(A)$ acts on the maximal forms $M^{4}$ as a scaling by a positive 
real number, i.e.\ they are orientation preserving. Therefore, $\Real(M^{4})$ 
may be split to the cones of positive and negative 
maximal forms  $\Real_{+}(M^{4})$ and $\Real_{-}(M^{4})$, preserved by $\Aut(A)$.

\begin{figure}[!ht]
\begin{center}
\includegraphics[height=\figuresize]{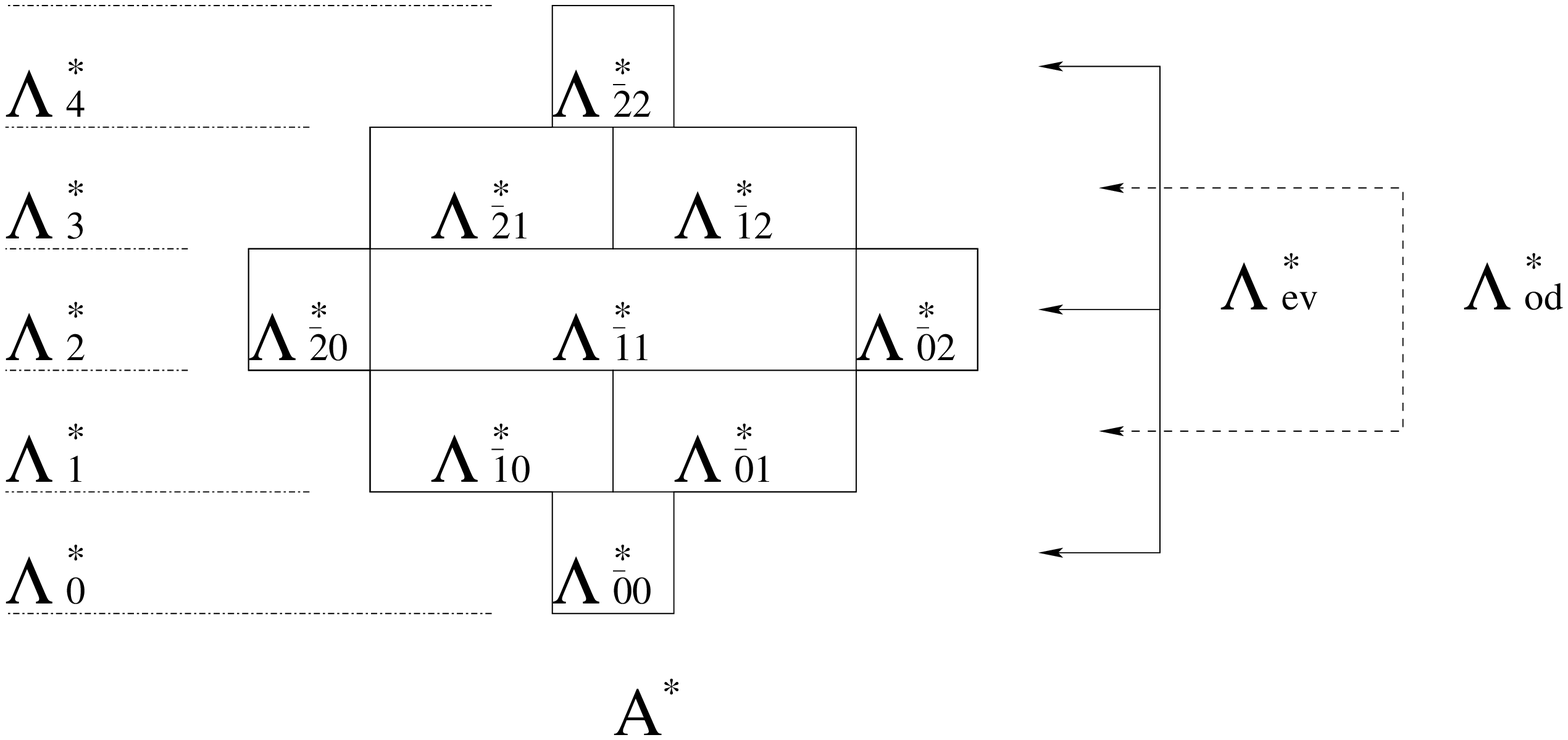}\hspace*{\figurehgap}\hspace*{\figurehgap}\includegraphics[height=\figuresize]{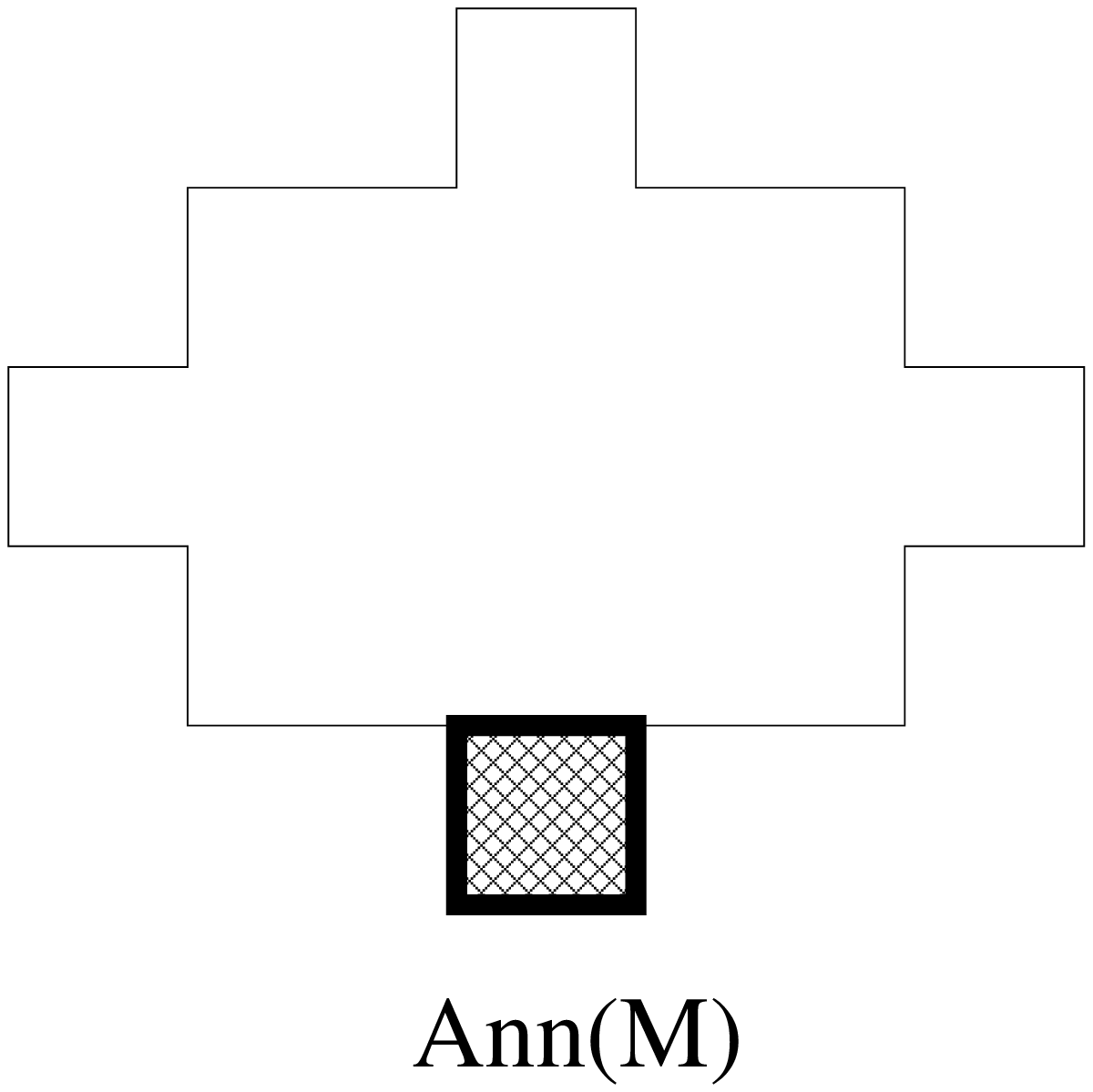}\\
\vspace*{\figurevgap}
\includegraphics[height=\figuresize]{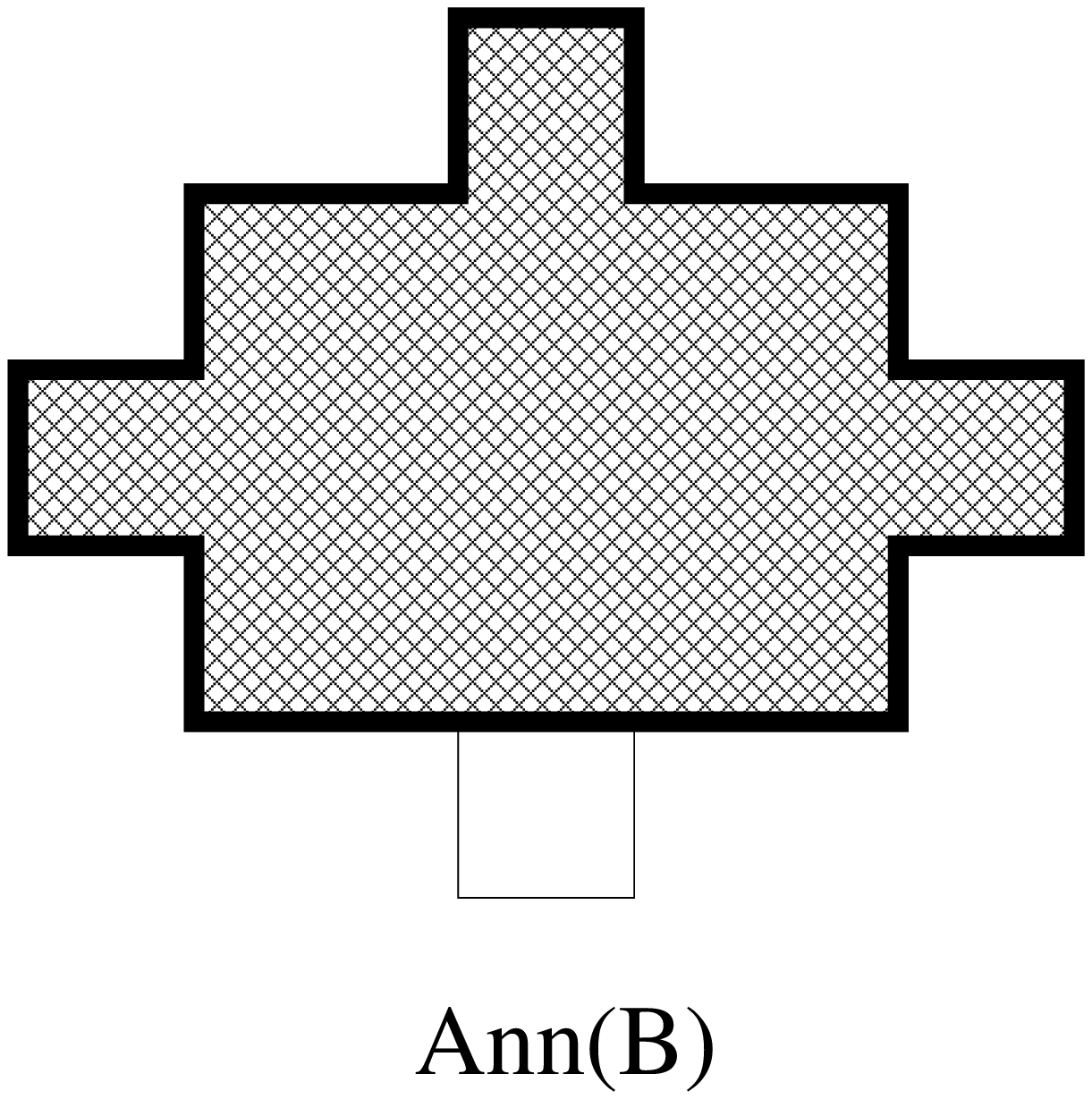}\hspace*{\figurehgap}\includegraphics[height=\figuresize]{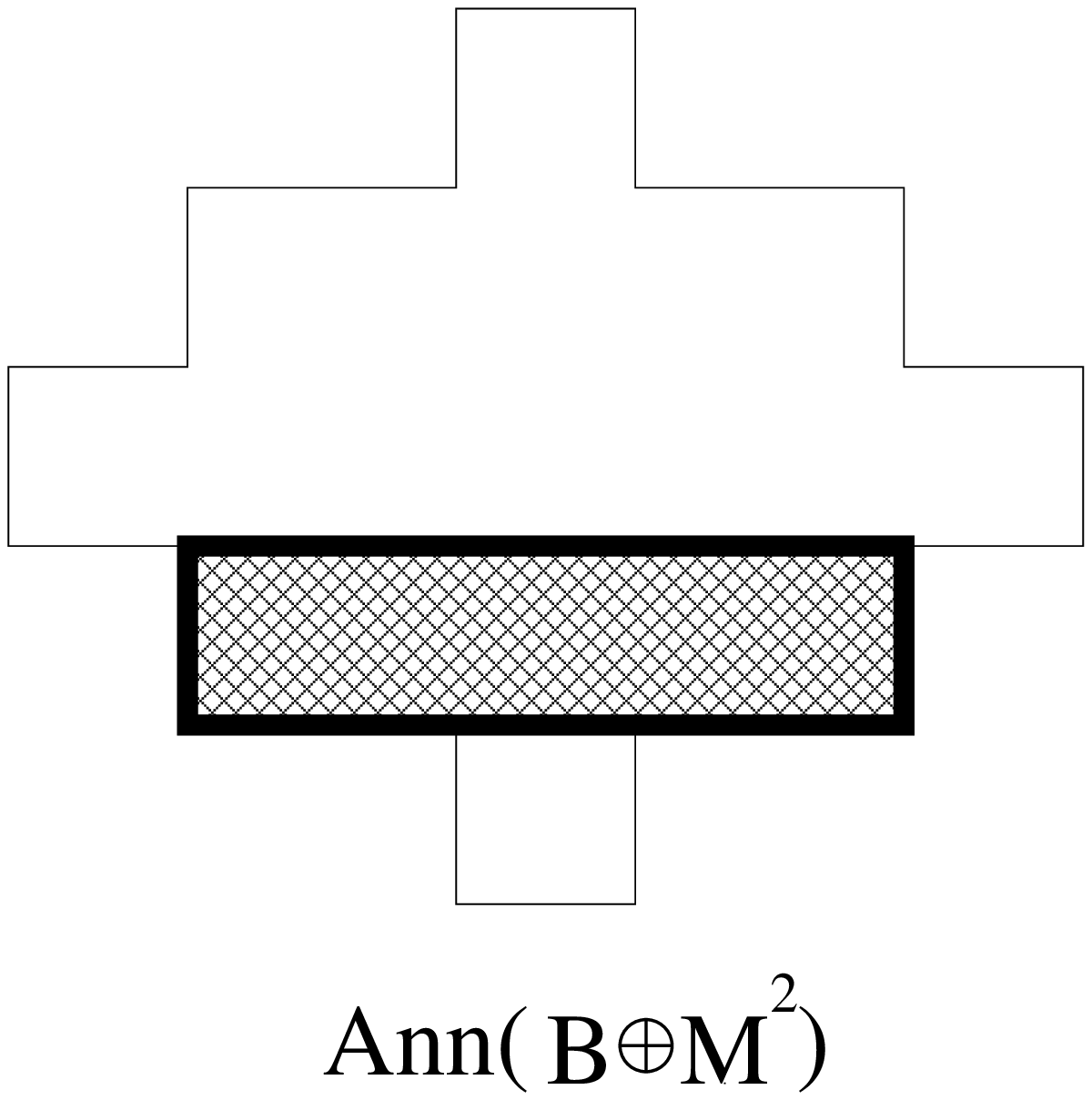}\hspace*{\figurehgap}\includegraphics[height=\figuresize]{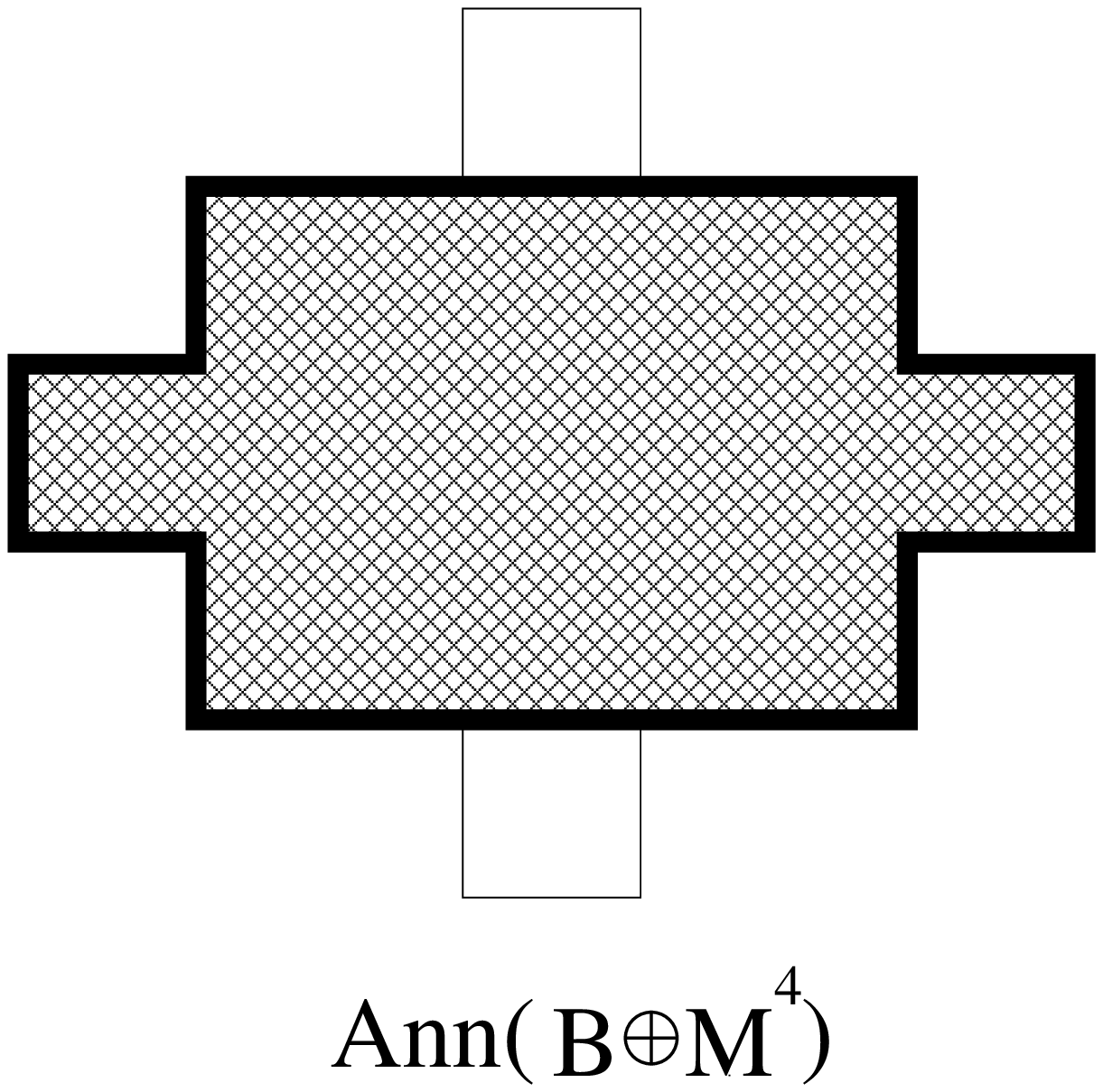}\\
\vspace*{\figurevgap}
\includegraphics[height=\figuresize]{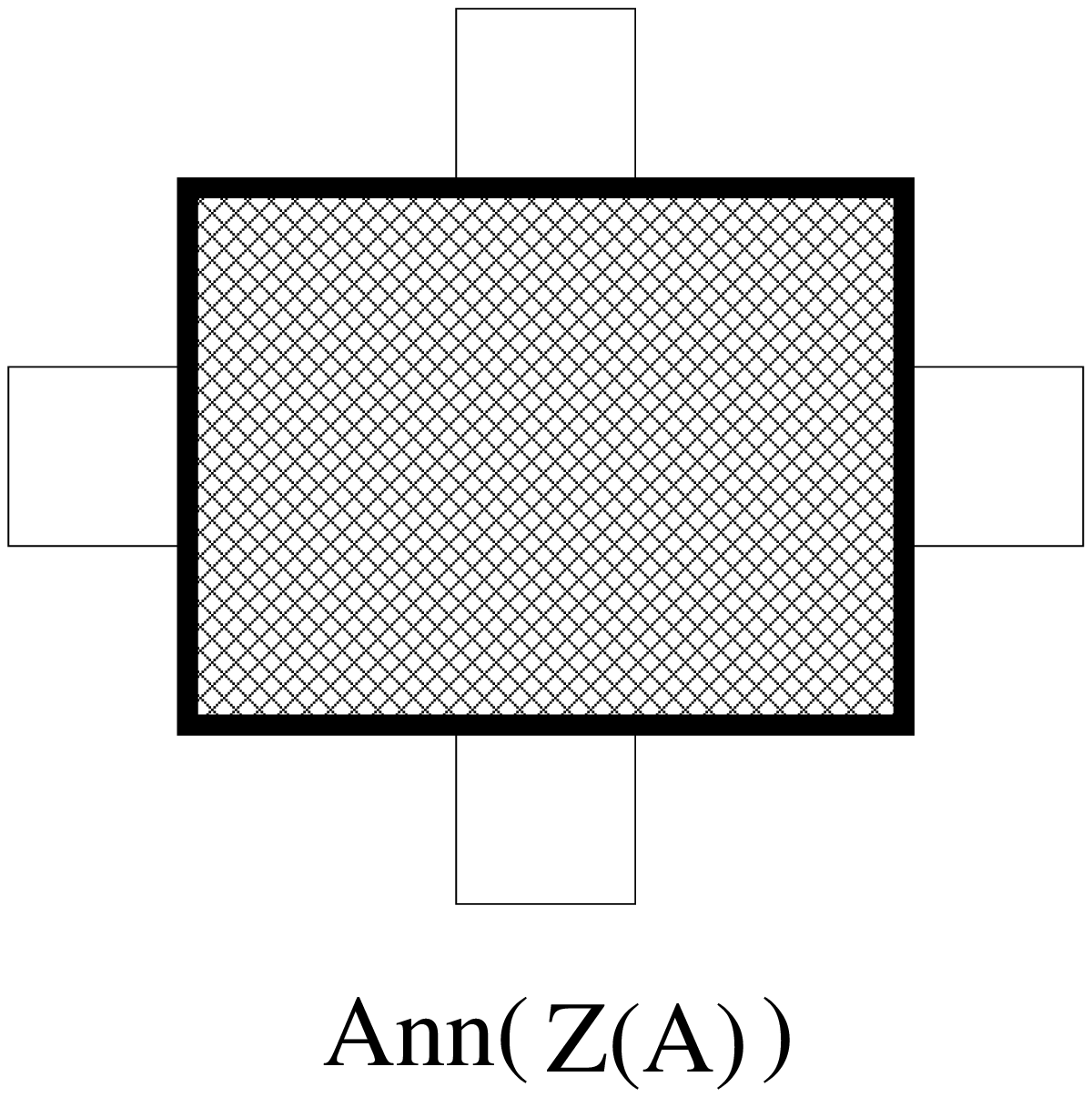}\hspace*{\figurehgap}\includegraphics[height=\figuresize]{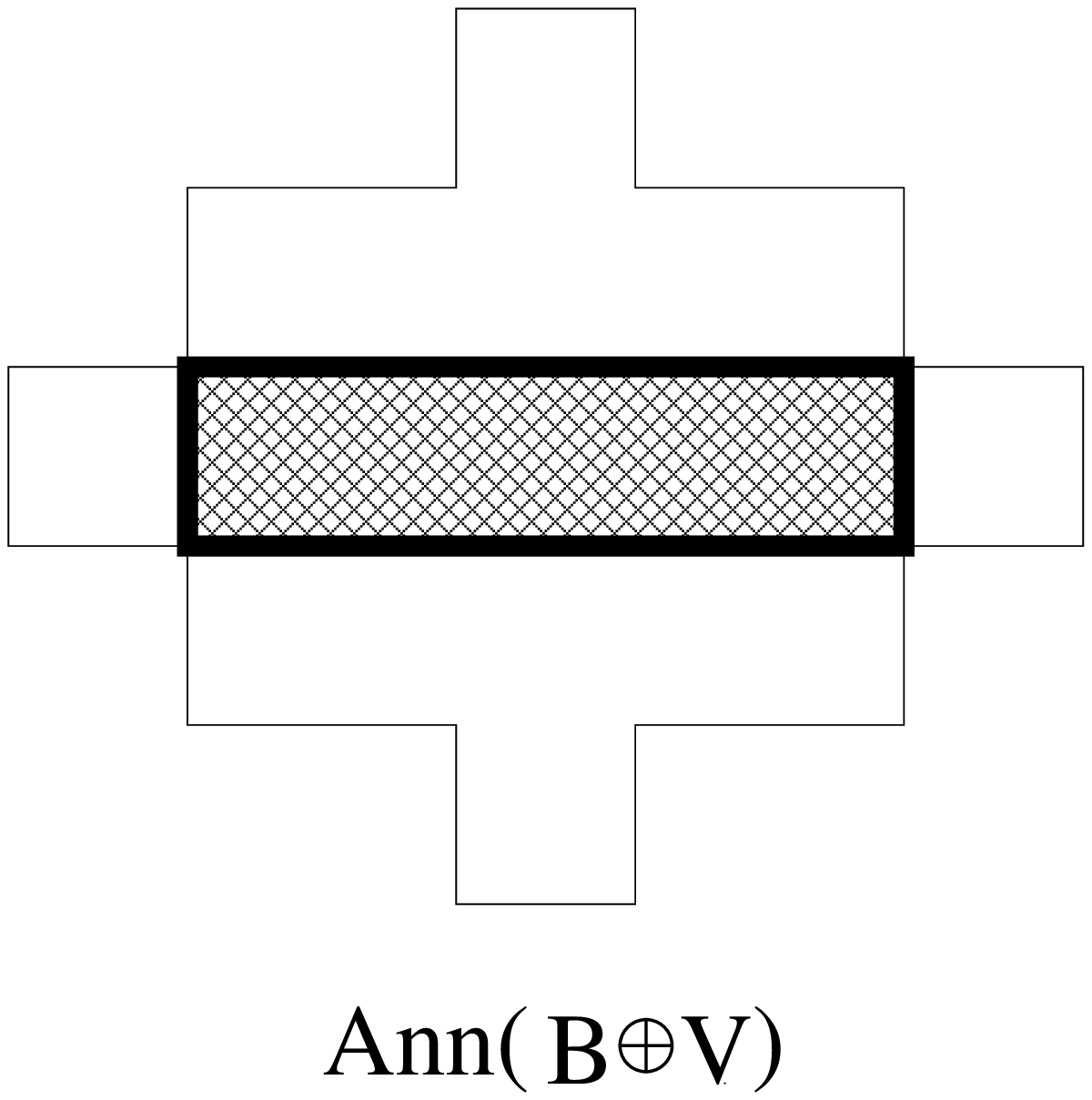}\hspace*{\figurehgap}\includegraphics[height=\figuresize]{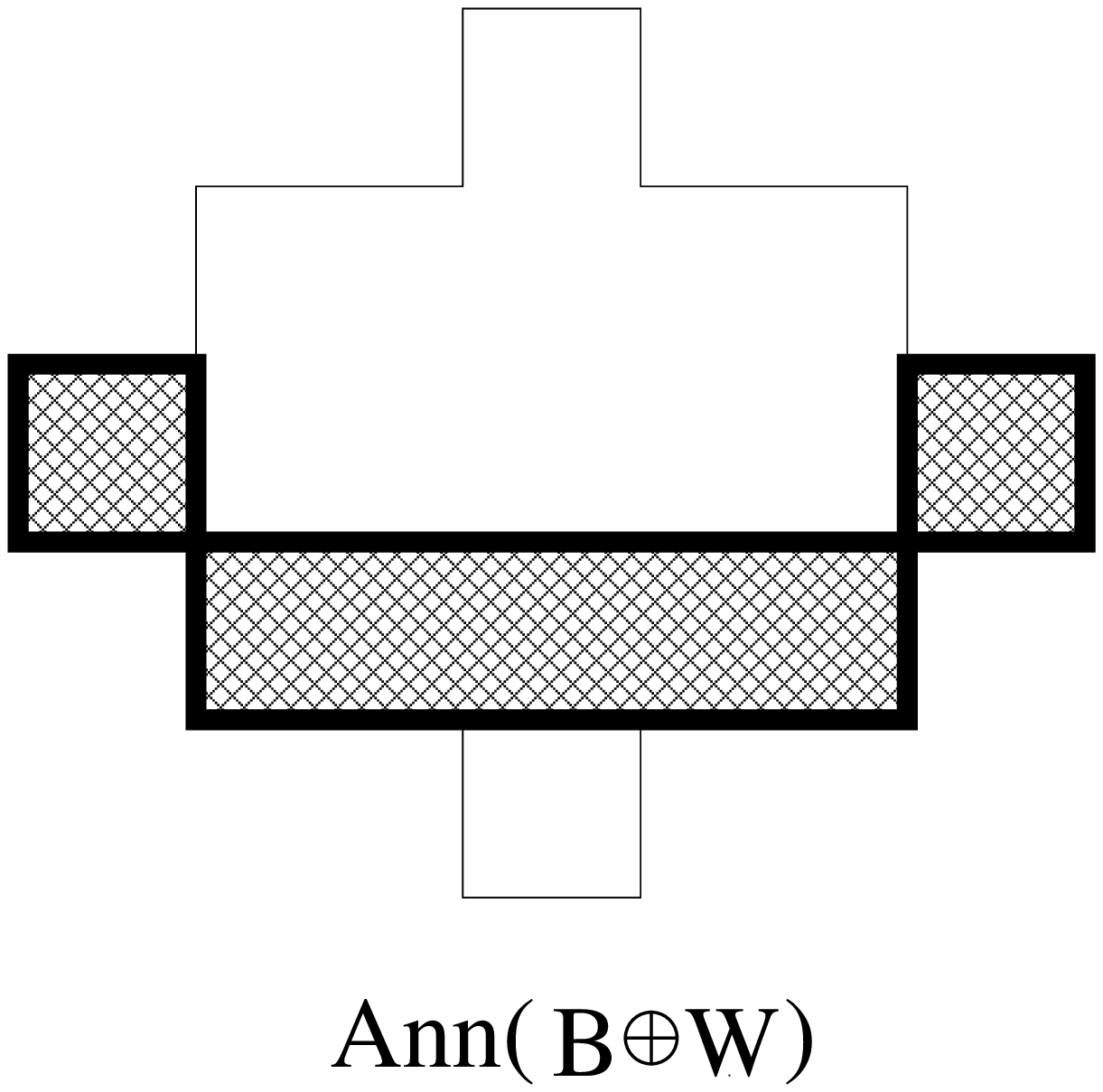}
\end{center}
\caption{Top left panel: illustration of $\Zs$, $\Z$ and $\Zt$-graded structure of the dual 
$A^{*}$ of a spin algebra $A$. Other panels: illustration of $\Aut(A)$-invariant 
indecomposable subspaces of $A^{*}$. One unit box depicts one complex dimension 
on all panels, shaded framed regions depict the pertinent $\Aut(A)$-invariant subspaces.}
\label{figspindualinvsubsp}
\end{figure}

\begin{Rem}
In order to proceed, we study the transpose action of $\Aut(A)$ on the dual vector 
space $A^{*}$ of $A$. From simple linear algebra, it automatically follows, that 
the indecomposable $\Aut(A)$-invariant subspaces of $A^{*}$ are:
$\Ann(M)$, $\Ann(B)$, $\Ann(B\oplus M^{l})$ ($l\in\{2,4\}$), $\Ann(Z(A))$, $\Ann(B\oplus V)$, $\Ann(B\oplus W)$, 
where $\Ann(\cdot)$ denotes the annulator subspace in $A^{*}$. 
These are illustrated in Fig.\ref{figspindualinvsubsp}.
\end{Rem}

We shall show that $\Aut(A)$ acts as the conformal Lorentz group on the four real dimensional 
$\Aut(A)$-invariant vector space $\Real(\Ann(B\oplus V))\equiv\Real(\Lambda_{\bar{1}1}^{*})$. For this, 
we use the fact that a spin algebra may be equipped with a Hopf algebra structure, given a chosen $\Zs$-grading. 
As usual in the Hopf algebra context\cite{lambe2002,cartier2006}, the unit is viewed as an $\eta:\,\C\rightarrow A$ linear map, 
and the algebraic product as an $\triangled:\,A\otimes A\rightarrow A$ linear map. 
The counit shall be an $\varepsilon:\,A\rightarrow\C$ linear map, and we set $\varepsilon:=b$. 
The antipode shall be an $S:\,A\rightarrow A$ linear map, and we set
$S(\cdot):=(\cdot)_{\ev}+(-1)(\cdot)_{\od}$. The swapping involution shall be 
an $\mathcal{I}:\,A\otimes A\rightarrow A\otimes A$ linear map, and we set to
$x\otimes y\mapsto\mathcal{I}(x\otimes y):=\sum_{p,q=0}^{2}\sum_{r,s=0}^{2}(-1)^{pr+qs}x_{\bar{p}q}\otimes y_{\bar{r}s}$ 
(for all $x,y\in A$). The coproduct shall be a $\triangle:\,A\rightarrow A\otimes A$ linear 
map, defined by the requirements $\triangle(\1):=\1\otimes \1$, 
$\triangle(x):=\1\otimes x+x\otimes \1$ (for all $x\in\Lambda_{\bar{1}0}\oplus\Lambda_{\bar{0}1}$), and 
that $\triangle$ is an $A\rightarrow A\otimes A$ algebra homomorphism, where 
$A\otimes A$ is equipped with the skew-natural product $(\triangled\otimes\triangled)\circ(I\otimes(J\circ\mathcal{I})\otimes I)$, 
$I$ being the $A\rightarrow A$ identity map and $J$ being the $A\otimes A\rightarrow A\otimes A$ swapping map. 
The holding of the Hopf relations on $(A,(\cdot)^{+},\eta,\triangled,\varepsilon,\triangle,S,\mathcal{I})$ 
may be checked by direct calculations. Given the fixed spin algebra part 
$(A,(\cdot)^{+},\eta,\triangled)$ it is seen that the coalgebra part 
$(\varepsilon,\triangle,S,\mathcal{I})$ is not conserved by all $\Aut(A)$ transformations, 
only under the $\Z$-grading preserving transformations $\Aut_{\Zs}(A)\rtimes\mathcal{J}\subset\Aut(A)$: 
the dressing transformations $N$ deform the part $(\triangle,S,\mathcal{I})$ 
to an other compatible coalgebra structure.

\begin{Rem} Equipped with the Hopf algebra notions, direct calculation shows 
that if we choose an $\omega\in\Real_{+}(M^{4})\setminus\{0\}$ positive maximal form 
element of $A$, the bilinear map
\begin{eqnarray}
G(\omega):\,\Real(\Ann(B\oplus V))\times\Real(\Ann(B\oplus V))\rightarrow\R,\cr
\qquad(a,b)\mapsto G(\omega)(a,b):=(a\otimes b|\triangle(\omega))
\end{eqnarray}
is a Lorentz signature metric, where $(\cdot|\cdot)$ denotes the duality pairing form. 
The action of $\Aut(A)$ preserves $G(\omega)$ up to a scaling factor, i.e.\ it acts 
on $\Real(\Ann(B\oplus V))$ as the conformal Lorentz group. The construction does 
not depend on the choice of the coproduct $\triangle$, i.e.\ does not depend on grading. The choice of $\omega$ 
is unique up to a scaling factor, and thus $G(\omega)$ is unique up to that as well.
\label{remmetric}
\end{Rem}

Next, we show that there are two charge-conjugate Dirac bispinor spaces embedded (not uniquely) in $A$, 
and thus the relation to the usual Clifford algebra formulation becomes clear. 
Given a coproduct $\triangle$, the four real dimensional $\Aut(A)$-invariant subspace 
$\Real(\Ann(B\oplus V))\subset A^{*}$ may be embedded into $\Real(\Lin(A))$ using the rule 
$s\mapsto (s\otimes I)\circ \triangle$ (for each $s\in \Real(\Ann(B\oplus V))$), 
which we call a \emph{\textbf{Pauli embedding}}. Given a four real dimensional vector 
space $T$ (modeling the spacetime tangent vectors at a spacetime point, or 
the momentum vectors of momentum space), a linear injection $T\rightarrow \Real(\Ann(B\oplus V))$ 
may be taken, which we call a \emph{\textbf{Pauli injection}}. The composition of a 
Pauli embedding and a Pauli injection shall be called \emph{\textbf{Pauli map}}, which is a 
$T\rightarrow\Lin(A)$ linear map. If we take such a Pauli map $\sigma$, in ordinary 
two-spinor calculus it corresponds to $\sigma_{a}^{A'A}$ in Penrose abstract indices\cite{penrose1984,wald1984}, 
where ${}_{a}$ corresponds to a $T^{*}$ index, while ${}^{A'}$ and ${}^{A}$ are spinor indices. 
Keeping these in mind, one can see that the object $g(\sigma,\omega)_{ab}:=b\,\sigma_{a}\sigma_{b}\,\omega$ 
is a Lorentz metric on $T$, because of Remark~\ref{remmetric}

\begin{Def} (Dirac adjoint, Dirac gamma map)
Let us fix a Pauli map $\sigma:\,T\rightarrow\Lin(A)$ and a positive maximal form $\omega$. 
The conjugate-linear map
\begin{eqnarray}
\tbar{(\cdot)}:\,A\rightarrow A^{*},\; x\mapsto \tbar{x}:=\frac{1}{2}g(\sigma,\omega)^{ab}\,b\sigma_{a}\Big(x^{+}\sigma_{b}(\cdot)+\sigma_{b}(x^{+})(\cdot)\Big)
\end{eqnarray}
is called the \textbf{Dirac adjoint}.
The linear map
\begin{eqnarray}
\gamma(\sigma,\omega):\,T\rightarrow\Lin(A),\; u\mapsto u^{a}\gamma(\sigma,\omega)_{a}(\cdot):=u^{a}\sqrt{2}\Big(\sigma_{a}(\cdot)+\sigma_{a}(\omega)(\cdot)\Big)
\end{eqnarray}
is called the \textbf{Dirac gamma map}.
\label{clifforddef}
\end{Def}

\begin{figure}[!ht]
\begin{center}
\includegraphics[height=\figuresize]{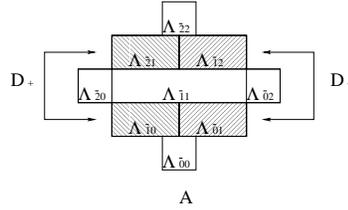}
\end{center}
\caption{Illustration of the embedding of the Dirac bispinor spaces 
$D_{+}$ and $D_{-}$ into the spin algebra $A$, given a Pauli mapping $\sigma$. 
One unit box depicts one complex dimension. Shaded regions indicate the subspaces 
$D_{+}$ and $D_{-}$, respectively.}
\label{figcliffordembedding}
\end{figure}

In order to see the properties of the Dirac adjoint and the Dirac gamma map, 
the following important subspaces of $A$ are introduced: let 
$D_{+}:=\Lambda_{\bar{1}0}\oplus\Lambda_{\bar{2}1}$ and 
$D_{-}:=\Lambda_{\bar{0}1}\oplus\Lambda_{\bar{1}2}$, where the $\Zs$-grading is 
understood to be the one subordinate to the Pauli map $\sigma$, or equivalently, 
the one used in the underlying coproduct. Clearly, these are four 
complex dimensional subspaces of $A$ with 
$D_{+} \cap D_{-} = \{0\}$ and $\left(D_{+}\right)^{+} = D_{-}$, $\left(D_{-}\right)^{+} = D_{+}$. 
Direct calculations show that $\tbar{(\cdot)}$ becomes non-degenerate and 
$\gamma(\sigma,\omega)$ satisfies the Clifford relations 
$\gamma(\sigma,\omega)_{a}\gamma(\sigma,\omega)_{b}+\gamma(\sigma,\omega)_{b}\gamma(\sigma,\omega)_{a}=2\,I\,g(\sigma,\omega)_{ab}$ 
over $D_{+}$ and $D_{-}$. These subspaces, however, are not preserved by the 
dressing transformations $N$, but are transformed to other similarly behaving embedded 
Dirac bispinor spaces in $A$. This embedding is illustrated in Fig.\ref{figcliffordembedding}.

Finally, the relation of $(\cdot)^{+}$-adjoining to the usual Hilbert scalar product adjoining is given, 
which is used for identification of antiparticle creation operators to particle annihilation 
operators in a traditional QFT setting. Let $u^{a}\in T$ be a future directed timelike or null vector in terms 
of $g(\sigma,\omega)_{ab}$, modeling a momentum vector on mass shell. Then, 
the bilinear form $A\times A\rightarrow \C,\,(x,y)\mapsto b\,(x^{+}y)+u^{a}\,b\,\sigma_{a}\,((x-\1\,b\,x)^{+}(y-\1\,b\,y))$ 
is a positive semidefinite $\Aut(A)$-invariant inner product. Given a coproduct $\triangle$, this induces 
a scalar product on $A$, and the adjoining operator $(\cdot)^{\dagger}$ with respect 
to that is the usual Hilbert adjoining identifying antiparticle creation operators 
and particle annihilation operators at a fixed momentum $u^{a}\in T$. Since 
it is momentum dependent (has a parameter $u^{a}$), and is not $\Aut(A)$-invariant, 
it is not practical to consider $(\cdot)^{\dagger}$ as basic ingredient of the internal degrees 
of freedom, encoded in the mathematical structure of $A$: it is rather constructed as a derived 
quantity, as shown. Also in this approach, the problematics of normal ordering does not appear: 
all polynomial expressions of $A$ are automatically normal ordered, by construction.

\section{Concluding remarks}
\label{concludingremarks}

A finite dimensional algebra was presented and its group of automorphisms 
was studied. The constructed algebra can physically be thought of as 
the creation operator algebra of a formal quantum field theory at fixed 
momentum of a spin 1/2 particle and its antiparticle. It was shown that the 
essential part of the pertinent automorphism group is basically the 
conformal Lorentz group. On the other hand, the remaining normal subgroup 
can be thought of as ``dressing transformations'' making ``dressed'' states 
from pure one-particle states. The proposed construction may be used in 
construction of quantum field theories on a non-perturbative basis: our 
approach is different than that of usual approach when the one-particle 
theory is defined, and then a Fock space of multiparticle states is built 
on top of that. In addition, a similar approach could be used in GUT 
attempts, as it provides a possibility to circumvent Coleman-Mandula theorem: 
the non-semisimpleness of the pertinent symmetry group could allow to have 
the spacetime and internal symmetries to be connected via the non-semisimple 
part, without introducing SUSY. An interesting idea would be to study 
these results in the light of \cite{furey2015}, which provides an 
$\mathrm{U}(1)\times \mathrm{SU}(3)$ unification attempt, also using an 
algebra automorphism group.

\section*{Acknowledgments}

I would like to thank M\'aty\'as Domokos and Rich\'ard Rim\'anyi for 
valuable discussion and expert's opinion on the algebra theory content 
of the paper, furthermore to P\'eter Vecserny\'es for ideas related to 
the Hopf algebra context. 
I would like to thank Tam\'as Matolcsi, P\'eter V\'an, 
Tam\'as F\"ul\"op and Dezs\H{o} Varga for motivations and discussion on the 
physical context of the presented idea. 
I would like to thank J\'ulia Ny\'iri for inviting the submission of this manuscript. 
This work was supported in part by the Momentum (``Lend\"ulet'') program of the 
Hungarian Academy of Sciences, as well as the J\'anos Bolyai Research 
Scholarship of the Hungarian Academy of Sciences.

\end{document}